\documentclass[twocolumn,showpacs,preprintnumbers,prl,floatfix]{revtex4}

\newcommand{\BE}{\begin{equation}}
\newcommand{\EE}{\end{equation}}
\newcommand{\BA}{\begin{eqnarray}}
\newcommand{\EA}{\end{eqnarray}}
\input epsf.tex
\usepackage{graphicx}% Include figure files
\usepackage{dcolumn}% Align table columns on decimal point
\usepackage{bm}% bold math
\begin{document}

\title{Analysis of microwave
propagation in arrays of dielectric cylinders}

\author{Bikash C. Gupta$^{1,2}$}\author{Zhen
Ye$^1$}\email{zhen@phy.ncu.edu.tw} \affiliation{$^1$Department of
Physics, National Central University, Chungli, Taiwan 32054\\
$^2$Department of Physics, University of Illinois, Chicago, IL,
USA 60612}

%\date{February 2, 2002}
\date{\today}

\begin{abstract}

We rigorously compute the propagation and scattering of microwaves
by regular and defected arrays of dielectric cylinders in a
uniform medium. Comparison with the previous experimental results
is made, yielding agreements in the transmission coefficients for
a certain range of frequencies. The results also show that
localized states are possible for defected arrays. However, the
experimentally claimed localized states are not observed.
Moreover, comparison is made with previous theoretical results.
The agreements and differences are pointed out.

\end{abstract}

 \maketitle

Waves surround us. Direct human communication is mainly conveyed
by acoustic waves, and is enriched by gestures which are passed
into our eyes through optical waves. Nowadays electronic waves are
also everywhere in our daily experiences such as audio \& video
systems, computers, and Nintendo games. Among many interesting
properties pertinent to waves, the phenomenon of wave localization
is probably most intriguing and remains as one of the unsolved
mysteries of last century.

The concept of wave localization was first proposed by Anderson
nearly half a century ago, initially for electronic systems
\cite{Anderson58}, then extended to classical waves including
microwaves to be addressed in this paper. It refers to situations
that purely due to multiple scattering of waves by disorders,
transmitted waves are confined or frozen in space and the wave
envelopes decay nearly exponentially along any
direction\cite{Lee}. Since its inception, wave localization has
attracted substantial efforts from all areas of expertise,
signified by the great body of literatures ranging from small
scales such as electronic systems, nano-structured materials, to
large scales such as seismic and ocean waves, as reviewed and
summarized in, for instance,
Refs.~\cite{Lee,Ishimaru,Meso,John,Ad,Datta,Nano,EA,loc,Seis,ocean}.
In spite of the efforts, however, some fundamental issues
concerning wave localization are still open \cite{Rama}.

A few prominent problems in the investigation lie in the
localization in two dimensional (2D) random media. First, although
it has been conjectured, thought to be genuinely valid, by the
scaling analysis\cite{gang4} that all waves are localized in 2D
random media for any given amount of disorders or impurities, the
experimental observation of 2D localization is scarce. Even for
the limited experimental results that are available \cite{McCall},
the disparity \cite{Meade,Ho} still remains in the interpretation
of the results. A main concern is whether the localization in 2D
has indeed been observed or not \cite{Meade}. Some authors further
pointed out that there still lacks a definite observation of 2D
wave localization, particularly for classical waves
\cite{Sigalas}. In addition, how waves are localized in 2D and
whether the localization features can be explained or fitted by
the existing theory \cite{DV} are among a few important issues yet
to be addressed.

It would be too ambitious to answer all the unsolved issues
dwelling on 2D localization in a single endeavor. However, a
further look at some of previous experimental observations of 2D
localization, in the hope of clarifying some confusions, seems
more realistic and imperative, and will therefore be the main
theme of the present paper.

Here we would like to concentrate on the experimental observation
of microwave propagation in 2D dielectric lattices, reported by
McCall {\it et al.} \cite{McCall}. The reason is not only because
the experiment is the first ever reported, but it has stimulated
so many later investigations, and its importance could be inferred
from the constantly increasing citation number. In \cite{McCall},
the authors measured transmission of microwaves through regular
and defected arrays of dielectric cylinders. The transmission
bands and photonic band gaps for the regular arrays are found to
be in excellent agreement with theoretical predictions. For the
defected arrays, the authors suggested the observation of a
localized state. This observation, however, has been contrarily
discussed and supported by two independent theoretical inspections
\cite{Meade,Ho}. In this paper, we will use the standard multiple
scattering approach first formulated by Twersky \cite{Twersky} to
further examine the experimental results. We will show that
localized states are indeed possible for the defected arrays.
However, the experimentally claimed localized states are not
observed. Comparison will also be made with previous theoretical
results. Agreements and differences are found and discussed. Some
contraries in the previous analysis are clarified.

The systems considered here are from Ref.~\cite{McCall} and they
are composed of almost ideal arrays of low-loss high dielectric
constant cylinders. The cylinders are placed in parallel in an
uniform medium to form a square lattice of lattice constant
denoted as d. The ratio of dielectric constants between the medium
and cylinders is 9. Two types of measurement are carried out in
\cite{McCall}. One is to measure the E-polarized microwave
transmission across the samples which are shaped as a rectangular
slab with thickness of 9d and width of 18d and the lattice
constant is 1.27 cm. The other is to measure the intensity
distribution at 11.2 GHz for a sample size of 7.6 cm x 7.6 cm with
a cylinder being removed from the center of the sample, i.~e.
defected array. However, the lattice arrangement for the second
measurement differs from the first in that the lattice constant is
1.59 cm; such an inconsistency in the lattice arrangements has
been discussed in \cite{Meade}. The authors in \cite{McCall}
suggested that the localized states are observed for the defected
array.

In our computation, we will take the physical parameters from
\cite{McCall}. However, we will consider larger lattice sizes when
appropriate to ensure the stability of the results. Both lattice
arrangements, i.~e. d = 1.27 and 1.59 cm corresponding to filling
factors 0.4487 and 0.2863 respectively, will be considered. The
cylinder radius is 0.48 cm. In addition to computing the
transmission across rectangular slabs, for comparison purpose we
also compute the transmitted intensity when the source in placed
inside a sample that takes a circular shape. The spatial
distribution of energy will also be calculated to compare with the
experiment. Moreover, the phase diagram method \cite{Emile} will
be used to investigate a coherence phenomenon associated with
localization. To be consistent with the experiment, we only
consider the E-polarized microwaves.

\begin{center}
\begin{figure}[hbt]
%\vspace{10pt}
\epsfxsize=2.5in\epsffile{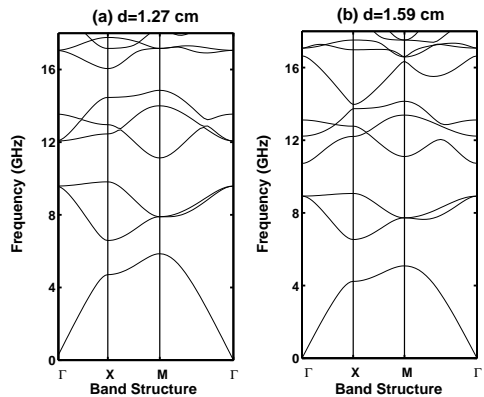} \caption{The band structures
of two 2D square lattices: (a) for d = 1.27 cm and (b) for d =
1.59 cm respectively. The dielectric constant is 9. The complete
band gaps are apparent for both lattices.} \label{fig1}
\end{figure}
\end{center}

Figure \ref{fig1} shows the band structures for the two 2D square
lattices aforementioned. The results on the left panel reproduce
that in \cite{McCall}, confirming our numerical codes. For later
discussion, the band structures of lattice constant 1.59 cm are
also plotted. Here we see that the two results are similar for
frequencies below the second gap. There are three prominent
complete bandgaps for d = 1.27, while there are only two complete
bandgaps for d = 1.59.

\begin{center}
\begin{figure}[hbt]
%\vspace{10pt}
\epsfxsize=2.5in\epsffile{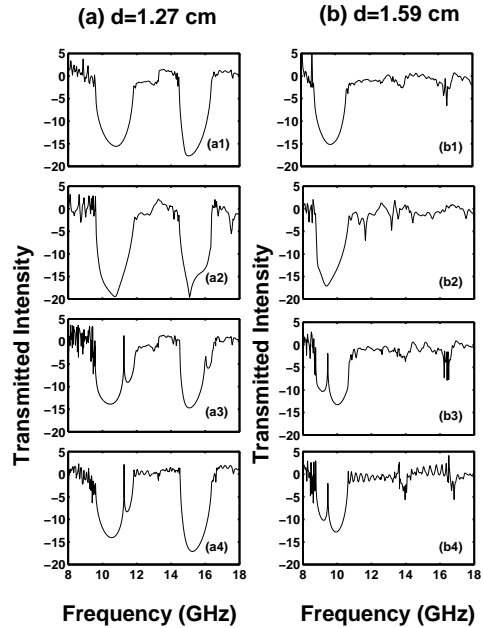} \caption{The transmitted
intensity is plotted as a function of frequency (in GHz) for (a) d
= 1.27 cm shown in the left panel and (b) for d = 1.59 cm shown in
the right panel respectively. (a1 \& b1): Regular system with size
of thickness 9d and width 18d; the wave propagates in the [10]
direction; both the source and the receiver are kept at one
lattice constant away from the both ends of the sample. (a2 \&
b2): The same as (a1 \& b1) except that the wave propagates in the
[11] direction. (a3 \& b3): System with a single defect, i.e., the
central cylinder is removed; the propagation is along the [10]
direction; both the source and the receiver are kept at one
lattice constant away from the both ends of the sample
respectively. (a4 \& b4): The source is moved into the center of
the sample with the central cylinder being removed; the sample
takes a circular shape of radius 9d.} \label{fig2}
\end{figure}
\end{center}

The transmission intensity is plotted as a function of frequencies
in Fig.~\ref{fig2}. The situations in (a1), (a2) and (a3) are
considered in \cite{McCall}. From this figure, we observe the
following. First for d = 1.27 cm: (1) The results for situations
considered by the experiment agree both qualitatively and
quantitatively well with the measurement for frequencies up to the
beginning of the third bandgaps at about 14 GHz; (2) While we did
not plot the case of propagation along the [21] direction, we plot
the case of the transmission from a source located inside the
sample, with a cylinder being deleted at the center (a4). The
result is in accordance with the case that the transmission is
measured across the sample. In both cases, there is a transmission
peak inside the valley at 11.27 GHz, slightly above the measured
value, which is outside the complete gap, but within the gap along
the [10] direction. The reason for the difference in the location
and magnitude of the peak from the experiment is due to the
resolution of plotting. (3) There are some significant differences
between the theory and experiment for frequencies starting from
the third gap. For these frequencies, we observe clearly a well
defined reduction in the transmission, well matching the band
structures in Fig.~\ref{fig1}, and the reduction is even deeper
than the second gap. The experimental results, however, show some
smeared reduction. Further calculations point to a few factors
contributing to the disagreement, such as disorders,
inhomogeneities in the radii of the cylinders, and the size of
samples.

The results of the transmission through a slab with d = 1.59 cm
are depicted on the right panel of Fig.~\ref{fig2}. These diagrams
show that the transmission is significantly prohibited in the
second gap, consistent with the band structure calculation. There
is also a peak in the reduction regime. The frequency at which the
energy distribution is measured is outside this forbidden regime.

An important feature in Figs.~(a3) and (a4) is the peak inside the
second gap along the [10] direction but outside the complete
bandgap, i.~e. within the pseudo gap. McCall et al. \cite{McCall}
attempted to explain this peak as the result of a localized state.
To support this view, McCall et al. further measured the energy
spatial-distribution at this peak and attempted, believed to be
successful, to show that there is indeed a localized wave at this
frequency. Unfortunately, this time they used the different
lattice with d = 1.59 cm. The inconsistency has been pointed out
in \cite{Meade}. Meade et al. \cite{Meade} argued that the peak is
due to the resonance of extended states, and the apparent
localized wave shown by Fig.~4 in \cite{McCall} is superficial and
is due to the resonance.

\begin{center}
\begin{figure}[hbt]
%\vspace{10pt}
\epsfxsize=2.5in\epsffile{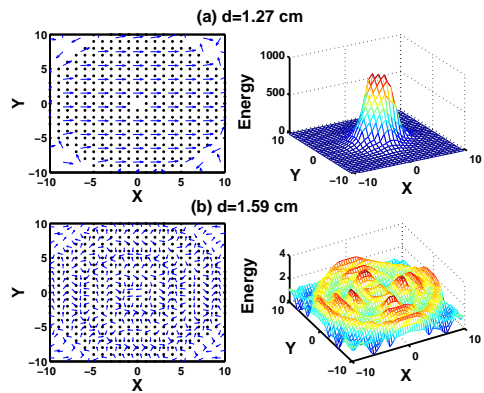} \caption{Single defect: Phase
and spatial energy distribution for frequency = 11.27 GHz with (a)
d = 1.27 cm and (b) d = 1.59 cm respectively. Note that at this
frequency there is a peak in the transmission for the system with
d = 1.27 cm (c.~f. Fig.~\ref{fig2}(a3, a4)), while there is no
such peak for d = 1.59 cm (c.~f. Fig.~\ref{fig2} (b3, b4)).
Hereafter, the trivial geometrical spreading effect is removed by
the normalization with regard to the energy directly from the
source. The black dots denote the cylinder.} \label{fig3}
\end{figure}
\end{center}

To resolve the disagreement between the measurement and the
previous theoretical analysis, we show in Fig.~\ref{fig3} the 2D
spatial distribution of energy at 11.27 GHz for the lattice of d =
1.27 cm and 1.59 cm with one cylinder being taken out at the
center of the sample. We also plot the associated phase diagram.
The source transmits at the center.

According to Refs.~\cite{Emile,Ye}, there will appear a coherence
for the phase of the localized wave. We brief this feature here.
The energy flow of EM waves is $\vec{J} \sim
\vec{E}\times\vec{H}$. By invoking the Maxwell equations to relate
the electrical and magnetic fields, we can derive that the time
averaged energy flow is $ <\vec{J}>_t \equiv \frac{1}{T}\int_0^T
dt \vec{J} \sim |\vec{E}|^2\nabla\theta,$ where the electrical
field is written as $\vec{E} = \vec{e}_E |\vec{E}|e^{i\theta}$,
with $\vec{e}_E$ denoting the direction, $|\vec{E}|$ and $\theta$
being the amplitude and the phase respectively. It is clear that
when $\theta$ is constant, at least by spatial domains, while
$|\vec{E}| \neq 0$, the flow would come to a stop and the energy
will be localized or stored in the space. The phase can be
represented by a vector defined as $\vec{v}(\vec{r}) =
\vec{e}_x\cos\theta + \vec{e}_y\sin\theta$, and the phase vectors
at various space points can be drawn on the 2D coordinates. The
coherence in the phases is a unique indication of localization.

Fig.~\ref{fig3}(a) shows that the wave is indeed localized at
11.27 GHz, as suggested in \cite{McCall} for the case of d = 1.27
cm. Here is shown that the energy is mostly confined within the
lattice and the prescribed phase coherence prevails. The
localization due to scattering by the cylinders is so strong that
the magnitude of the intensity surpasses the energy directly from
the source. However, this localized wave is not what was observed
in the experiment. With the experimental setup of d = 1.59 cm, the
experimentally claimed localization is actually absent, as
suggested by Fig.~\ref{fig3} (b) where the phase vectors point to
different directions indicating the flow of energy. Nevertheless,
the ripples in the energy distribution mimics that observed in the
experiment. We note that for all cases, the sample size considered
here is larger than that in the experiment.

\begin{center}
\begin{figure}[hbt]
%\vspace{10pt}
\epsfxsize=2.5in\epsffile{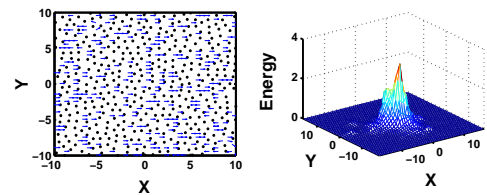} \caption{The completely random
array: The phase and energy diagram for frequency = 11.27 GHz. The
black dots refer to the cylinders.} \label{fig4}
\end{figure}
\end{center}

To further confirm that localization is indeed possible at
11.27GHz, we also considered the case of a complete random array
which is otherwise regular with d = 1.27 cm. The results are shown
in Fig.~\ref{fig4}. Here the wave localization indeed appears. The
energy is localized at the center of sample with phase vectors
aligning along one direction. Due to the randomness, the peak
value of the localized energy is reduced significantly.

\begin{center}
\begin{figure}[hbt]
%\vspace{10pt}
\epsfxsize=2.5in\epsffile{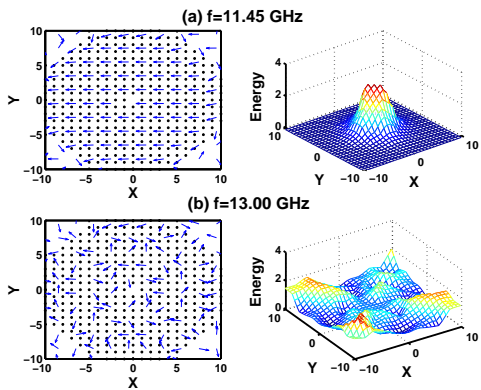} \caption{Single defect for
d=1.27 cm: Phase and energy diagram for (a) frequency = 11.45 GHz
which is slightly above 11.27 GHz at which we observed a peak in
the transmission in Fig.~\ref{fig2}, and for (b) frequency = 13
GHz which lies between the second and third gaps (see
Fig.~\ref{fig1}(a)).} \label{fig5}
\end{figure}
\end{center}

We further consider two other frequencies for the case of regular
array d = 1.27 cm with one cylinder being removed: one is at 11.45
GHz, slightly above 11.27 GHz but still within the pseudo gap
along the [10] direction, and the other one lies at 13 GHz between
the second and third gaps. The results are illustrated in
Fig.~\ref{fig5}. They show that the localization is apparent for
11.45 GHz while absent for 13 GHz. It is interesting to notice
that the pointing direction of the phase vectors is mostly
reversed from the case with the lower frequency, comparing to
Fig.~\ref{fig3} (a). It is also worth to noting that the
localization shown by both Figs.~\ref{fig3} and \ref{fig5}  occurs
outside the complete band gap. These results will eventually give
insight to the relation between gap and localization. The absence
of localization in defected arrays may be due to the finite size
of samples, in the view of previous studies. But it might also
hint that waves are not necessarily always localized in 2D, as
suggested before \cite{Emile}.

\begin{center}
\begin{figure}[hbt]
%\vspace{10pt}
\epsfxsize=2in\epsffile{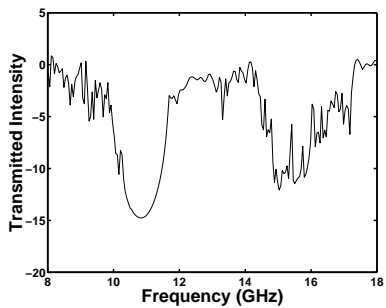} \caption{Transmitted intensity
as a function of frequency for a disordered array of size 9 x 18.
The waves propagates along the [10] direction. The randomness is
twelve percent of random displacement of the cylinders within a
circle around their original regular positions with regard to the
lattice constant.} \label{fig6}
\end{figure}
\end{center}

Finally, to examine the difference between the theoretical and
experimental transmission results when comparing the left panel of
Fig.~\ref{fig2} and Fig.~3 in \cite{McCall}, we investigate the
effect of disorders. For brevity, we only show the effect of the
positional disorder which is introduced by moving the cylinders
around their regular lattice position to a certain extent. Here
the disorder is measured as 12\% of random displacement of the
cylinders within a circle around their original regular positions
with regard to the lattice constant. The result is shown in
Fig.~\ref{fig6}. The result indeed shows that the disorder tends
to degrade the transmission reduction in the third gap to a
certain degree, while the disorder effect is not so significant
for the second gap.

In summary, in this paper we have shown a numerical analysis of
the previous experimental results on microwave propagation through
arrays of dielectric cylinders embedded in a uniform medium
\cite{Meade} by the standard multiple scattering method.
Comparison has been also made with other theoretical results. The
results reveal that localized states are possible for defected
arrays. However, the experimentally claimed localized states are
not observed. The disorder effects are also presented. The present
work is hopped to stimulate further experimental observations.

Thanks to NSC and NCU for supports.


\begin{thebibliography}{99}

\bibitem{Anderson58} P. W. Anderson, Phys. Rev. {\bf 109}, 1492 (1958).

\bibitem{Lee} P. A. Lee and T. V. Ramakrishnan, Rev. Mod. Phys. {\bf
57}, 287 (1985).

\bibitem{Ishimaru} A. Ishimaru, {\it Wave propagation and scattering in
random media}, (Academic Press, New York, 1978), Vols. I and II.

\bibitem{Meso} {\it Scattering and localization of classical waves in random
media}, edited by P. Sheng (World Scientific, Singapore, 1990).

\bibitem{John} S. John, Phys. Today {\bf 40}, 32 (1991).

\bibitem{Ad} A. Lagendijk and B. A. van Tiggelen, Phys. Rep. {\bf 270}, 143
(1996).

\bibitem{Datta} S. Datta, {\it Electronic transport in mesoscopic
systems} (Cambridge University Press, New York, 1997).

\bibitem{Nano} V. A. Markel and T. F. Thomas, {\it Optics of
nanostructured materials}, (Wiley \& Sons Inc., New York, 2001)

\bibitem{EA} E. Abrahams, S. Kravchenko, and M. P. Sarachik, Rev.
Mod. Phys. {\bf 73}, 251 (2001).

\bibitem{loc} M. Janssen, {\it Fluctuations and localization}
(World Scientific, Singapore, 2001); and references therein.

\bibitem{Seis} R. Hennino, {\it et. al.}, Phys. Rev. Lett. {\bf 86},  3447 (2001).

\bibitem{ocean} G. I. Barenbatt, M. E. Vinogradov, and S. V.
Petrovskii, Oceanology, {\bf 35}, 202 (1995).

\bibitem{Rama} T. V. Ramakrishnan, Pramana - J. Phys. {\bf 56}, No. 2, 149 (2002).

\bibitem{gang4} E. Abrahams, P. W. Anderson, D. C. Licciardello, and
T. V. Ramakrishnan, Phys. Rev. Lett. {\bf 42}, 673 (1979).

\bibitem{McCall} S. L. McCall, P. M. Platzman, R. Dalichaouch, D. Smith, and S. Schultz,
Phys. Rev. Lett. {\bf 67}, 2017 (1991).

\bibitem{Meade} R. D. Meade, A. M. Rapper, K. D. Brommer, J. D.
Joannopoulos, and O. L. Alerhand, Phys. Rev. {\bf B} 48, 8434
(1993).

\bibitem{Ho} M. Sigalas, C. M. Soukoulis, E. N. Economou, C. T.
Chan, and K. M. Ho, Phys. Rev. {\bf B}, 48, 14121 (1993).

\bibitem{Sigalas} M. M. Sigalas, C. M. Soukoulis, C.-T. Chan, and
D. Turner, Phys. Rev. B{\bf 53}, 8340 (1996).

\bibitem{DV}  D. Vollhardt and P. W\"olfle, Phys. Rev. B {\bf 22},

\bibitem{Twersky} V. Twersky, J. Acoust. Soc. Am. {\bf 24}, 42
(1951).

%\bibitem{Chen} Y.-Y. Chen and Z. Ye, Phys. Rev. Lett. {\bf 87},
% 184301 (2001).

\bibitem{Emile} E. Hoskinson and Z. Ye, Phys. Rev. Lett. {\bf 83}, 2734
(1999).

\bibitem{Ye} Z. Ye, S. Li, and X. Sun, Phys. Rev. E 66, 045602(R)
(2002).

\end{thebibliography}
\end{document}